\documentstyle[12pt]{article}

\def\GeV{\,{\rm GeV}}
\def\TeV{\,{\rm TeV}}

\def\keV{\,{\rm keV}}
\def\MeV{\,{\rm MeV}}
\def\sec{\,{\rm sec}}

\def\rcm{\,{\rm cm}}

\def\Mpc{\,{\rm Mpc}}

\def\eV{{\,\rm eV}}

\def\cmm2{{\,\rm cm^{-2}}}
\def\cm2{{\,{\rm cm}^2}}
\def\cmm3{{\,{\rm cm}^{-3}}}
\def\gcmm3{{\,{\rm g\,cm^{-3}}}}

\def\mpl{{m_{\rm Pl}}}
\def\la{\mathrel{\mathpalette\fun <}}
\def\ga{\mathrel{\mathpalette\fun >}}
\def\fun#1#2{\lower3.6pt\vbox{\baselineskip0pt\lineskip.9pt
  \ialign{$\mathsurround=0pt#1\hfil##\hfil$\crcr#2\crcr\sim\crcr}}}
\def\km{\rm \,km}
\def\yrs{\rm \,yrs}

\begin{document}
\baselineskip=14pt
\begin{titlepage}
\vspace*{-62pt}
\begin{flushright}
FERMILAB-Conf-92/382-A \\
January 1993
\end{flushright}
\vspace{0.5in}
\centerline{\large \bf DARK MATTER:  THEORETICAL PERSPECTIVES}
\footnotetext{Presented at the NAS Special Colloquium on
Physical Cosmology, Irvine, March 1992; to appear
in the {\it Proceedings of the National Academy of Sciences.}}

\vspace{.5in}
\centerline{Michael S.~Turner}
\vspace{.3in}

\begin{center}

{\it Departments of Physics and of Astronomy \& Astrophysics\\
Enrico Fermi Institute, The University of Chicago, Chicago, IL~~60637-1433}\\
\vspace{.2in}
{\it NASA/Fermilab Astrophysics Center\\
Fermi National Accelerator Laboratory,
Batavia, IL~~60510-0500}\\

\end{center}

\end{titlepage}

\newpage
\vspace{5cm}

\centerline{\bf Abstract}
\begin{quotation}

\noindent I both review and make the case for the current theoretical
prejudice:  a flat Universe whose dominant constituent is
nonbaryonic dark matter, {\it emphasizing that this is still a prejudice
and not yet fact.}  The theoretical motivation for nonbaryonic
dark matter is discussed in the context of current elementary-particle
theory, stressing that:  (i) there are no dark matter candidates
within the standard model of particle physics;
(ii) there are several compelling candidates within
attractive extensions of the standard model of particle
physics; and (iii) the motivation for these
compelling candidates comes first and foremost
from particle physics.  The dark-matter problem is now a pressing
issue in both cosmology and particle physics, and the
detection of particle dark matter would provide evidence
for ``new physics.''  The compelling candidates are:
a very light axion ($10^{-6}\eV -10^{-4}\eV$);
a light neutrino ($20\eV -90\eV$); and a heavy neutralino ($10\GeV
- 2\TeV$).  The production of these particles in the
early Universe and the prospects for their detection are
also discussed.  I briefly mention more exotic
possibilities for the dark matter, including a nonzero
cosmological constant, superheavy magnetic monopoles,
and decaying neutrinos.

\end{quotation}

\newpage
\pagestyle{plain}
\setcounter{page}{1}
\newpage

\section{Overview}

One of the simplest yet most fundamental questions
we can ask in cosmology concerns the quantity
and composition of the matter in the Universe:
What is mass density, expressed as
a fraction of the critical density, $\Omega_0$, and
what are the contributions of the various constituents, e.g., baryons,
photons, and whatever else?  (The critical density
$\rho_{\rm CRIT} = 3H_0^2/8\pi G =
1.88h^2\times 10^{-29}\gcmm3 =
1.05\times 10^4\eV\cmm3$ where $H_0 =100h\,{\rm km\,sec^{-1}
\,Mpc^{-1}}$.)  The answer to this question
bears upon almost every topic discussed at this Colloquium:
the expansion age and fate of the Universe; the origin of
structure in the Universe and CBR anisotropies;
galactic disks, rotation curves, and morphology;
cluster dynamics; gravitational lensing; and
the distribution of light and mass.  The only thing we know
with great precision is the contribution of photons,
$\Omega_\gamma = 2.49 h^{-2} \times 10^{-4}$ (assuming
$T_{\gamma 0}=2.73\,$K), and neutrinos,
$\Omega_\nu = 1.70 h^{-2}\times 10^{-4}$ (assuming
all three species are massless); and based on primordial nucleosynthesis,
we know the contribution of baryons to within
a factor of two, $\Omega_Bh^2 =0.01-0.02$ \cite{bbn}.

In principle, the classic kinematic tests---luminosity-red shift,
angular size-red shift, number count-red
shift, and so on---can be used to
determine $\Omega_0$ (provided that we know the equation
of state of the Universe) \cite{Sandage}.
To date these tests have not been successful because they require
standard objects of one sort or another (luminosity, size, or
number density), though hope was expressed at this
Colloquium that new techniques
may change this situation (e.g., K-band Hubble diagram,
K-band number counts, type I or II supernovae, and so on).  At present,
our knowledge of $\Omega_0$ derives primarily from dynamical estimates
that sample small, often atypical environments (e.g., rich
clusters and bright spiral galaxies).
There is an exception, the recent attempts to infer
$\Omega_0$ based upon the peculiar motion of the Local Group,
which interestingly enough yield a value
for $\Omega_0$ of order unity and with small error
estimates \cite{rr,yahil}.  Beyond the fact that
this measurement supports theoretical prejudice, it may well
come the closest to weighing a large, fair sample of the Universe.

What is clear is that most of the mass density is accounted for by
dark matter (i.e., matter that emits nor absorbs
any radiation) and that $\Omega_0$ is at least 0.1---and
perhaps as high as order unity.  Since primordial nucleosynthesis
provides very convincing evidence that baryonic matter
can contribute no more than 10\% of critical density \cite{bbn},
we are left with two possibilities:  (i) conclude that
$\Omega_0$ lies at its lower bound, that $\Omega_B$ lies at its
upper boundary, and that $h\la 0.5$, in which case
$\Omega_0 \simeq \Omega_B \simeq 0.1$; or (ii)
conclude that there is a ``gap'' between $\Omega_0$ and
$\Omega_B$ and consider the consequences.

While the second possibility is the more radical, the
evidence for a gap, though not yet conclusive, continues to mount.
If we accept this gap as real, and make the leap all the way to
a flat Universe there are important implications:
By a wide margin most the Universe is comprised
of nonbaryonic matter, and
because there are no nonbaryonic dark-matter
candidates within the standard model of the elementary
particles, the dark-matter problem becomes one of pressing
interest in particle physics also.  Particle physics
rises to the occasion:  In several of the most attractive
extensions of ``their standard model'' there are hypothetical
particles, whose motivations are unrelated to cosmology,
but whose relic abundance is close to the closure density.
The most promising are:  an axion of mass $10^{-6}\eV-
10^{-4}\eV$; a neutralino of mass $10\GeV - 2\TeV$;
and a neutrino of mass $90h^2 \eV$.\footnote{A
massive neutrino is not considered part of the standard model
because neutrino masses are not accommodated within
standard model.}

Most theorists would agree that a flat Universe
dominated by nonbaryonic matter is the
most attractive hypothesis, so attractive it is sometimes
forgotten that it is still just that.
This paradigm has become an
almost indispensable crutch for those who study the formation of
of structure.  In fact, I know of no viable model
of structure formation based upon a Universe with
$\Omega_0 = \Omega_B \simeq 0.1$.\footnote{Peebles'
isocurvature baryon model comes close, but as I
understand it, the model requires that $\Omega_B
\sim 0.2$ and $h\sim 0.8$ \cite{pib}.}

That being the case,
it is important that we take our theoretical beliefs
seriously enough to test them!  At our
disposal are a host of laboratory experiments
and observational tests.
They include cosmological measurements of $\Omega_0$,
$H_0$, the age of the Universe,
CBR anisotropies, large-scale structure and
so on.  In the laboratory there are efforts
to directly detect halo dark-matter particles,
to produce new particles at high-energy
accelerators, to detect dark-matter annihilation
products (coming from the sun or the halo),  as well as
a multitude of experiments that search for evidence
for neutrino masses.

\section{Weighing the Universe}

Measuring the mean density of the Universe is no simple
task; nor is summarizing the measurements and putting
them in perspective \cite{mass}.  Simply
put, one would like to weigh a representative volume
of the Universe, say $100h^{-1}\Mpc$ on a side.  Easier
said than done.  Because of the inconclusiveness of the
kinematic methods, I will focus on the dynamical measurements.

The dynamical measurements probe the mean density
in a less than ideal way:  A dynamical measurement,
e.g., the virial mass of a cluster, is converted
into a mass-to-light ratio which, when multiplied
by the mean luminosity density (which itself has
to be determined), yields an estimate of
the mean mass density.\footnote{In the $B_T$ system
the critical mass-to-light ratio is $1200h$, in solar
units.}  There is an obvious drawback:
One has to {\it assume}
the mass-to-light ratio derived for the object, or
portion thereof, is ``typical'' of the Universe as a whole.
With that as a preface---and a warning---let me proceed.

Mass-to-light ratios derived for the solar neighborhood
are very small, of order unity, and taken as a universal
mass-to-light ratio imply a value of $\Omega_0$ of much less
than 1\%.  Using instead the mass-to-light ratio
inferred from the inner luminous regions of spiral
and elliptical galaxies, of order ten or so, one infers
a value for $\Omega_0$ of somewhat less than 1\%.  Based upon
this evidence, most would agree that luminous matter
contributes less than 1\% of critical density \cite{fg}.

The flat rotation curves of spiral galaxies give
strong evidence that most of the mass in spiral
galaxies exists in the form of dark halos; assuming that
the halo material is distributed with spherical symmetry
(for which there is only minimal evidence), the density
of the halo dark matter decreases as $r^{-2}$ \cite{flat}.  Many would
cite the flat rotation curves of spiral galaxies as the
strongest evidence that most of the material in the
Universe is dark.  Using the mass-to-light ratios
derived from the flat rotation curves of spiral galaxies
one infers values of $\Omega_0$ in the range of 3\% to
10\%.  Since there is presently no convincing evidence
for a rotation curve that falls as $r^{-1/2}$, indicating
convergence of the total mass of the galaxy, one should
regard these estimates as {\it lower limits} to $\Omega_0$
(again, based upon this technique).

There is some evidence for dark matter in elliptical
galaxies and even dwarf galaxies, though it is much
harder to come by, as one must measure velocity dispersions
rather than rotation curves \cite{elliptical}.

The oldest evidence for dark matter, dating back
to the work of Zwicky \cite{zwicky}, involves clusters;
simply put, there isn't nearly enough mass associated with
the light to hold clusters together.  The masses of clusters
are derived using the virial theorem and involve certain assumptions:
the distribution of galactic orbits must be specified
and the clusters must be assumed to be ``well relaxed.''
The values for $\Omega_0$ deduced from
cluster mass-to-light ratios range from 10\% to 30\%,
though we should be mindful of the underlying assumptions
(current observations seem to indicate that clusters are
not well relaxed)
and the fact that any material that is distributed
spherically symmetrically outside the region where
galaxies reside would not contribute to the virial
masses derived.  And of course, the
fundamental assumption is that cluster
mass-to-light ratios are typical, though less than 1 in 10
galaxies resides in a cluster.   We should note too that
dark is a relative term:  It is now known that much, if
not the majority, of the baryonic mass in clusters exists
in the form of hot, x-ray emitting gas, that is ``dark''
to an optical telescope \cite{xray}.

The virial masses of small groups and binary galaxies
also provide evidence for dark matter, though the problem
of interlopers is a severe one.  The gravitational arcs
produced by the lensing effect of clusters also indicate
the presence of cluster dark matter.  {\it Evidence for dark
matter in the Universe is nowhere lacking.}

In my biased and very brief summary I have
saved the best for last, a measurement that comes close to
weighing a representative sample of the Universe of
order $100h^{-1}\Mpc$ on a side.  It involves tying
our well measured velocity with respect to the CBR,
about $620\km \sec^{-1}$, to the inhomogeneous distribution
of matter in the nearby Universe.  In effect, it is
a simple problem in Newtonian physics:  requiring that
our velocity be produced by the inhomogeneous distribution
of galaxies allows us to weigh a very large sample of
the Universe.  Here too assumptions are made:
that the distribution of galaxies traces the mass
at some level and that the bulk of our peculiar velocity
arises from galaxies inside the survey volume and
not outside.  Using the red shift survey based
upon the IRAS 1.2 Jy catalogue, two groups have
inferred values of $\Omega_0$ that are close to unity:
$\Omega_0 \simeq b^{1.7}$ with statistical errors
of order 0.3 \cite{rr,yahil}.
Here $b\equiv (\delta n_{\rm GAL}/n_{\rm GAL})/
(\delta \rho /\rho )$ is the so-called bias factor,
that in the simplest way accounts for the fact
that bright galaxies may not faithfully trace
the mass distribution.  (I should mention that attempts
to reconstruct the local density field from the
measured peculiar velocity field also leads to
a large value for $\Omega_0$ \cite{yahil}.)

To summarize the summary:

\begin{itemize}

\item Luminous matter (in the
form of stars and associated material) provides
at most 1\% of the critical density.

\item  The flat rotation curves of spiral galaxies and virial
masses of clusters indicate that the bulk of the
mass density in the Universe is dark.

\item The dark matter is less condensed than the luminous
matter (as evidenced by galactic halos).

\item $\Omega_0$ is at least 0.1, and the bulk of the data
are consistent with $\Omega_0 =0.2\pm 0.1$ ($\pm 0.1$ is {\it
not} might to be a statistical error flag).

\item  Primordial nucleosynthesis constrains the fraction
of critical density contributed by baryons to be between
1\% and 10\% (more precisely, $0.01 \la \Omega_B h^2 \la 0.02$).

\item  There is growing evidence for a gap between
$\Omega_B$ and $\Omega_0$.

\end{itemize}

A minimalist view is that we have a consistent
solution:  $\Omega_B = \Omega_0 \simeq 0.1$ and $h\la 0.5$.
The grander---and more radical---view
is that there is a gap between $\Omega_B$ and $\Omega_0$, that $\Omega_0 =1$,
and that we live in a Universe dominated by
nonbaryonic dark matter.  From a theoretical perspective
this is the most attractive scenario---and it may even be true!

Three points before we go on; as many have emphasized
it may well be that there are several kinds of dark matter \cite{carr}.
Unless $h\ga 1$ primordial nucleosynthesis already indicates
evidence for dark baryons; moreover, baryons could in principle
account for all the dark matter in galactic halos and possibly
even clusters (provided $h\la 0.5$).  Dark baryons could exist
in the form of black holes, neutron stars, or very low mass stars.
Three large-scale efforts are well under way to search for
dark matter in the form of low-mass stars in the halo of our
galaxy using their microlensing of stars in the LMC \cite{micro}.

While black holes may appear to the ideal dark-matter candidate,
there are not.   Black holes formed in the
contemporary Universe ultimately trace their origins to
baryons, and thus can contribute no more than about 10\% of
critical.  While it is possible that mini black holes, holes
much less massive than a solar mass, were produced in the
early Universe from the primeval plasma and could today
provide the critical density, a plausible mechanism
for producing the right number without other deleterious
consequences (e.g., black hole evaporations today
producing gamma rays) is lacking \cite{pbh}.

If $\Omega_0 =1$, then the question arises as to where the bulk of the
matter is, as most dynamical measurements indicate $\Omega_0 \sim 0.1-0.3$.
This is the $\Omega$-problem.  It could be that
galactic halos are very large and that clusters sit at the
center of gigantic distributions of dark matter, or that much of the
material exists in low-luminosity galaxies (so-called biasing),
or even that it exists in a form of smoothly distributed
energy density, e.g., relativistic particles or a cosmological
constant.  In that regard one of the very nice features
of neutrino dark matter is that neutrinos, owing to their
large velocities, would likely remain smooth on scales out
to several $\Mpc$.  In any case we know that the dark matter
is less condensed than luminous matter, indicating that
it does not have the ability to dissipate energy.  This
means that it could be in the form of particles that interact
very weakly, or tied up in large objects made of baryons
(e.g., dead stars or dwarfs).

\section{The Evidence for a Flat Universe!}

Before pursuing the hypothesis of a flat Universe dominated
by nonbaryonic dark matter let me quickly
summarize the evidence in support of it.

\begin{itemize}

\item There is evidence for a gap between $\Omega_B$ and
$\Omega_0$.

\item A dynamical explanation for our own peculiar velocity
seems to indicate that $\Omega_0$ is close to unity.

\item Some kinematic measurements of $\Omega_0$ based
upon galaxy counts indicate that $\Omega_0$ is close to unity \cite{counts}.

\item Structure formation in a low-$\Omega_0$ Universe is more
difficult and requires larger amplitude density perturbations and
may not be consistent with the smoothness the CBR \cite{sf}.

\item One of the most attractive scenarios of the early Universe,
inflation, unambiguously predicts a flat Universe \cite{guth}.

\item The Dicke-Peebles timing argument \cite{dicke}:
If the Universe is
not flat, then we must conclude that we live at a special
time when the curvature terms and matter density terms are comparable.

\end{itemize}

Needless to say the evidence is not overwhelming; it does,
however, make a case for taking the hypothesis of a flat
Universe dominated by nonbaryonic dark matter seriously.

\section{Nonbaryonic Dark-matter}

If we adopt $\Omega_0 =1$, then the gap between
$\Omega_0$ and $\Omega_B$ is significant and necessitates
that a new form of matter be the dominant constituent
of the Universe.  The point of this section
is to emphasize that particle
physicists too were pushed to nonbaryonic dark matter for
reasons solely based upon particle physics:
As a consequence of addressing very fundamental problems in particle
physics, the existence of new particles was predicted,
particles as it turned out whose relic cosmic
abundance was close to the critical density.  This could
just be a coincidence, or it could be an important hint
that we are on the right track.

\subsection{The standard model of particle physics}

Over the past two decades particle physicists have
constructed a fundamental theory that accounts for
all known phenomena at energies below about $300\GeV$
(down to length scales of order $10^{-16}\rcm$).
They call it ``the standard model'' \cite{sm}; mathematically,
it is a nonAbelian gauge theory based upon the group
$SU(3)_C\otimes SU(2)_L\otimes U(1)_Y$.  The
$SU(3)_C$ part, known as quantum chromodynamics,
describes the strong interactions (the interactions
that bind quarks in hadrons).\footnote{The interactions
between hadrons, e.g., between neutrons and protons,
which use to be referred to as the strong interactions,
are now believed to be analogous to van der Waals forces,
here residual forces between color
neutral objects, and hence not fundamental \cite{waals}.}
The $SU(2)_L\otimes U(1)_Y$ part
describes the electroweak interactions.  An important
part of the standard model is the notion that the
electromagnetic and weak interactions are not separate
phenomena, rather different aspects of the unified
electroweak force.

The fundamental particles of the standard models are
three families of quarks and leptons ($u$, $d$,
$c$, $t$, and $b$ quarks and $\nu_e$, $e^-$, $\nu_\mu$,
$\mu^-$, $\nu_\tau$, and $\tau^-$ leptons), and
12 gauge bosons (8 Gluons, $W^+$, $W^-$, $Z^0$, and
the photon) that mediate the fundamental interactions.
All the gauge bosons have been seen; the top quark
remains to be discovered; and there is only indirect---but
very strong indirect---evidence for the
existence of the tau neutrino.
All the particles participate in the electroweak
interactions; only quarks carry color and
participate in the strong interactions.

While the 8 Gluons and the photon are massless,
the $W^\pm$ and $Z^0$ bosons are not; this reflects
the least well understood aspect of the standard:
symmetry breaking.  The full symmetry
of the electroweak interactions is hidden; the simplest
explanation is the Higgs mechanism and involves a new
class of fundamental (scalar) particles:  Higgs
bosons, which have not yet been seen.  Hidden symmetry is analogous to
the magnetization of a ferromagnet:  at low temperatures,
due to spin interactions the
state of the ferromagnet with lowest free energy
is characterized by aligned spins and a net magnetization,
and thus does not exhibit rotational invariance.
The ground state of the Higgs field at low temperatures, due to its
self interactions, breaks the
symmetry of the electroweak interactions and
in so doing makes the $W^\pm$ and $Z^0$ bosons
massive (and accounts for the masses of the quarks
and leptons as well).  The aspects of the standard
model involving the gauge particles and quarks
and leptons have been tested to very high precision
(in many cases to better than 1\%); there is no
direct evidence for the Higgs mechanism, and it is
possible that something else accounts for
the hidden symmetry.  One of
the primary goals for the SSC is the elucidation
of symmetry breaking, e.g., by the
production of Higgs bosons.

The standard model is a neat little package; in accounting
for all ``known particle physics'' it also
explains the absence of other phenomena.  For example,
why are neutrinos so light (or perhaps massless)?
The $SU(2)_L$ symmetry forbids a mass for the neutrinos
(in the absence of righthanded neutrinos).
Why is the proton stable (or at least very long-lived)?
Again, in the standard model
it is not possible to have proton decay without
violating other symmetries of the standard model.\footnote{This statement
is true at the classical level; subtle quantum
effects associated with instantons and the like
lead to baryon-number violation.  At temperatures
$\ga 200\GeV$ these processes are probably very important
and may play a role in explaining the origin of the baryon
asymmetry of the Universe \cite{sphaleron}.}  Similar
considerations forbid interactions that violate lepton number.

\subsection{New physics beyond the standard model}

The tapestry of the standard model is not without
loose threads.  Like the standard cosmology it
has shortcomings that point to something
grander; they include:

\begin{itemize}

\item Quantization of charge:  quarks and leptons are
separate families of particles, yet the charges of the
quarks are to high precision an integer multiple of one-third
the charge of an electron.

\item A related issue:  why are there two kinds of matter particles
(quarks and leptons) and three families
of quarks and leptons?  Are quarks and leptons fundamental, or
are they made of ``smaller'' entities?

\item In the standard model the fundamental forces are
``patched'' together, rather than truly unified.

\item The standard model has more than 20 ``input parameters''
($\sin^2\theta_W$, quark and lepton masses, mixing angles, etc.)
that must be specified.

\item Disparity of scales:  the scale of the weak interaction
$G_F^{-1/2}\sim 300\GeV$ is much, much less than that of
gravity, $G^{-1/2}\sim 10^{19}\GeV$ (the ``hierarchy problem'').

\item A related issue:  how to keep the Higgs light enough
to break the electroweak interaction at a scale of $300\GeV$
in the face of quantum corrections that should drive
its mass to the highest energy scale in the theory ($10^{19}\GeV$).

\item The strong $CP$-problem:  within the standard model
quantum effects (instantons again) lead to $CP$ violation
in the strong interactions and should lead to an electric-dipole
moment for the neutron that is $10^9$ times larger than the
current upper limit.

\item Where and how does gravity fit in?

\end{itemize}

These considerations lead most particle physicists
to believe that there must
be a ``grander'' theory.  Moreover, the mathematical
tools at hand---nonAbelian gauge theories,
supersymmetry, superstrings, to mention three---allow
very attractive and powerful theoretical speculations
that address all of these issues.  These
speculations lead to the prediction of
new particles, some of which are stable (due to new
conservation laws) or are at least long-lived (due
to their small masses and/or very weak interactions).
Further---and this is the cosmological bonus---some
of these new, long-lived particles
have relic abundances that are comparable to the critical
density.  This didn't have to be; the relic
abundance of a particle species
is determined by its mass and interactions.
{\it This is either the big hint or
the grand misdirection.}

To put things in perspective here
is a very brief summary of the extensions of the standard
model and the dark-matter candidates they predict.

\begin{itemize}

\item Peccei-Quinn symmetry \cite{PQ}:  this is a very minimal
extension of the standard model designed to solve
the strong-$CP$ problem.  It is considered
by many to be the best solution and
automatically arises in many supersymmetry
and superstring models.  Another consequence
of PQ symmetry is the existence of a very long-lived,
light (pseudoscalar) particle---the axion---which is a
prime dark-matter candidate.

\item Majoron models \cite{majoron}:
these are modest
extensions of the standard model designed to accommodate neutrino
mass, and thereby allow the three ordinary neutrino
species to be dark matter candidates.

\item Supersymmetry \cite{susy}:  low-energy supersymmetry
is perhaps the most well studied extension of the standard
model.  Supersymmetry, the symmetry that relates bosons
and fermions, dictates that for every fermion there
be a bosonic partner (and vice versa)---thereby doubling
the particle content of the standard model.  First and foremost,
supersymmetry addresses
the hierarchy problem, ``stabilizing'' the
mass of the Higgs boson and putting scalar particles on
a firm footing.  It also paves the way for
the unification of gravity (when supersymmetry is gauged
it leads to general relativity).  Supersymmetry
must be a broken symmetry since the known particles do
not have equal-mass partners; the superpartner masses
are generically expected to be of order $10\GeV$ to $1000\GeV$
(of order the electroweak scale).  In almost all models
the lightest superpartner (or LSP) is stable and is a linear
combination of the photino and higgsino, known as
the neutralino.  The neutralino is a prime dark matter candidate.

\item Technicolor \cite{TC}:  is a very attractive idea for
replacing the Higgs mechanism with a mechanism akin
to the BCS mechanism in the BCS
theory of superconductivity.  A stronger
version of QCD---technicolor---leads to the formation
of bound states of techniquarks, and these bound
states play the role of the
Higgs.  Technicolor addresses the hierarchy problem as
the mass of the Higgs is set by the energy scale at
which technicolor becomes ``strong'' (just as the
mass of the hadrons is set by the scale at which
color becomes strong) and eliminates the need for
scalar particles.   However, it is an attractive
idea that has been very difficult to implement:  There is
currently no viable model of technicolor.
Whether or not it predicts the existence of
dark-matter candidates remains to be seen.

\item Grand unification \cite{GUT}:  the basic goal of grand unified
theories (GUTs) is to truly unify the strong, weak,
and electromagnetic interactions within a single gauge
group with one coupling constant.  The simplest
GUT is based upon the gauge group $SU(5)$ and predicts a proton
lifetime of $10^{30}\yrs$, which, sadly, has been falsified.  Other
GUTs include $SO(10)$, $E6$, $E8$, and on and on.
That unification is even possible---given that the coupling
strengths of the different interactions are so different
at low energies---is remarkable.  In nonAbelian gauge
theories coupling strengths vary (or ``run'')
with energy (logarithmically); the strengths of the three
known interactions seem to become equal at an energy scale of about
$10^{16}\GeV$ or so, which sets the scale of grand
unification.\footnote{About a decade ago the convergence
of the coupling constants occurred in ordinary GUTs at
an energy scale of about $10^{14}\GeV$; better measurements
of $\sin^2\theta_W$ indicate that such a convergence
does not occur in nonSUSY GUTs, but does in SUSY GUTs
at an energy scale of about $10^{16}\GeV$.}
Among other things, GUTs predict
proton decay, neutrino masses, and the existence of
superheavy magnetic monopoles (masses of order the
unification scale)---the last two being dark-matter
candidates.  In many GUTs neutrino masses arise via
the ``see-saw mechanism'' \cite{seesaw} and
$m_\nu \simeq m_l^2/{\cal M}$ where $m_l$ is the charged lepton
mass, and ${\cal M}$ is an energy scale associated with unification
(not necessarily the unification scale itself---perhaps
orders of magnitude smaller).
This explains why neutrino masses are so very small---and
in many models suggests that neutrinos may have masses
in ``the $\eV$ range'' (anywhere from $\mu{\rm eV}$ to
tens of $\eV$).

\item Superstrings \cite{ss}:  superstring theories unify all the
forces (including gravity) in a finite quantum theory (WOW!)
and are most naturally formulated in ten dimensions
(suggesting the existence of six extra spatial dimensions
that today must be compactified).
The fundamental objects of the theory are 1-dimensional string-like
entities whose size is order $10^{-33}\rcm$.  The expectations
for the superstring are high:  ultimately,
explanations for everything---quark/lepton
masses, coupling constants, the strong-$CP$ problem, the
number of families, spartner masses, the electroweak
scale.  The path has been more difficult
than expected, and there have been few definite predictions
(that are not wrong).  Broadly, superstring theory
provides theoretical support for the axion, supersymmetry,
grand unification, and neutrino masses---providing
motivation for all the dark-matter candidates mentioned above.

\end{itemize}

Of course, there are other ideas that I have not
mentioned because at present they do not seem
viable.  For example, preons, which were postulated
as the constituents of quarks and leptons, and
higher-dimensional analogs of superstrings, known as membranes.

\subsection{Two birds with one stone}

Particle dark matter is attractive
because new particles that owe their existence to attempts
to solve very fundamental puzzles in particle physics have
a relic abundance of order the
critical density!  Historically, such coincidences
have been a sign that one is on the right track.\footnote{For
a while, some believed that one could get three birds with
one stone:  Cosmions, dark-matter particles of mass $4\GeV$
to $10\GeV$ with scattering cross sections of order $10^{-35}\rcm^2$,
were proposed to solve both the solar-neutrino and the
dark-matter problems.  This possibility is all but ruled
out on both theoretical---the corresponding annihilation
cross section leads to a cosmion abundance that is too
small in both the sun and the cosmos---and experimental grounds---cosmions
should have been detected in dark-matter searches \cite{cosmion}.}

While there are now literally
dozens of particle dark matter candidates,
there are but a handful of particles whose existence
owes to well motivated attempts to solve important
problems in particle physics and whose relic
abundance is in the right ballpark.  They are:

\begin{itemize}

\item The neutralino \cite{neutralino}.  In most supersymmetry models,
the neutralino is the lightest supersymmetric partner
and is stable (due to a new symmetry called $R$-parity).
Its interactions with ordinary matter are roughly the
strength of the weak interactions, and this fact
ultimately explains why its relic abundance is of
order the critical density.  At present, supersymmetric
models have many parameters that must be dialed in,
and the mass of the neutralino is only known to
be somewhere between $10\GeV$ and $2\TeV$.

\item The axion \cite{axion}.  Peccei-Quinn symmetry seems to be
the best solution to the nagging strong-$CP$ problem.
The mass of the axion depends upon a single parameter:
the energy scale of PQ symmetry breaking, $f_{\rm PQ}$,
and $m_a\sim m_\pi^2 /f_{\rm PQ} \sim
10^{-5}\eV\,(10^{12}\GeV /f_{\rm PQ})$.
The strength of the axion's couplings to ordinary matter
are proportional to its mass.  When the axion was
first invented, only one scale of symmetry breaking
was known:  the weak scale and there seemed to be
a unique prediction for its mass, around $200\keV$.
This idea was quickly falsified.  It is now realized
that there are likely to many energy scales in Nature, the GUT scale,
the Planck scale, the intermediate scale and so on.
The symmetry-breaking scale has
been constrained, largely
by astrophysical and cosmological arguments, to
lie in the interval, $10^{10}\GeV \la f_{\rm PQ} \la
10^{13}\GeV$, corresponding to an axion mass in
the range $10^{-6}\eV$ to $10^{-3}\eV$ \cite{axlimits}.  This also
happens to be the range where the relic abundance
of axions is of order the critical density.

\item Light neutrino.  The neutrino exists;
it comes in three varieties; and we know its
relic abundance to three significant figures,
$113\cmm3$ per species.  Further, essentially
all extensions of the standard model predict
that neutrinos have mass, and the see-saw mechanism
implies masses in the general range of $\eV$,
give or take a factor of $10^3$ or so.

\item Dark horses.  There are also a few well
motivated long shots.  They include the superheavy
magnetic monopole:  It is a generic prediction
of GUTs; the only problem is its abundance,
without inflation far too many monopoles are
produced, and with inflation essentially
no monopoles are produced \cite{mono}.
There is the supersymmetric
partner of the axion, the axino, which arises
in theories with both PQ symmetry and supersymmetry \cite{axino}.
Its mass is expected to be in the $\keV$ range,
and its abundance is significantly less than
neutrinos as it decouples much earlier.

\end{itemize}

\subsection{Why not baryons or modified gravity?}

The particle dark matter hypothesis is a radical
solution; are there other alternatives that are
less radical or perhaps more attractive?
I think not, but to convince the reader let me mention two such
ideas:  $\Omega_B \sim 1$
and modified Newtonian dynamics.

Primordial nucleosynthesis provides the
best determination of the amount of
baryonic matter in the Universe, pinning
down the number density of baryons to within
a factor of two.  To be sure, the arguments
involve assumptions about the Universe in the
distant past.  Over the years many have suggested
alternative scenarios of primordial nucleosynthesis
that would allow one to evade the nucleosynthesis
and have $\Omega_B \sim 1$ \cite{malaney}.
The most recent attempt involved the role of
large inhomogeneities that
might have been produced in the quark/hadron transition
it it were strongly first order.  It was hoped
that such inhomogeneities would allow $\Omega_B\sim 1$.
This possibility is now ``doubly forbidden.''
As discussed at this Colloquium inhomogeneous nucleosynthesis
allows very little, if any, loosening of the standard
bound \cite{bbn};
moreover, numerical simulations of the quark/hadron transition
suggest that such inhomogeneities would not have
arisen in the first place, as the transition is
at best a weakly first-order phase transition, and perhaps
not a phase transition at all (more like recombination).

Theorists are rarely criticized for their conservatism!
Moreover, it seems that every theorist worth his
salt has tried to find a theory of gravity to supplant
Einstein's.  So one might have expected that
theorists would have
embraced Milgrom's modified Newtonian dynamics (MOND)
\cite{milgrom}.
The basic idea of MOND is that the form
of Newton's second law is modified
for accelerations less than about $cH_0\sim 10^{-7}
\rcm\sec^{-2}$, $F \simeq m a^2/cH_0$, thereby eliminating
the need for dark matter to explain flat rotation
curves.  While theorists are more than ready to
consider modifications to Einstein's theory, especially
in light of superstring theory, to most theorists MOND
looks like a nonstarter.
The reason is simple:  it is purely a Newtonian theory,
and attempts to formulate it in terms of a relativistic
field theory have been unsuccessful.  Without such
a formulation one cannot construct a cosmological
model or evaluate its predictions for
the many tests we have of relativistic theories of
gravity---bending of starlight, precession of the perihelion of Mercury,
gravitational red shift, radar time delay, and
the myriad of tests offered by the binary pulsar.
If that were not bad enough, it has been argued that
MOND can be falsified on
the basis of rotation curves measured for galaxies
of very different sizes \cite{nomond}.

In sum, theorists have looked hard for other explanations;
I believe that it is fair to say that the particle
dark matter explanation is the most attractive.  Whether or not
it proves to be correct is another matter.

\section{Dark-matter Relics:  Origins}

Since an important motivation for particle dark matter
is the fact that the relic abundance of these handful
of promising candidates is comparable
to the critical density it is worth reviewing how a
cosmological relic arises.   There are several
qualitatively different mechanisms for particle
dark-matter production in the early Universe.

\subsection{Thermal relics:  hot, warm, and cold}

Much---but not all---of the history of the Universe is
characterized by thermal equilibrium.  So long as
equilibrium pertains the abundance of a massive particle
relative to photons is:\footnote{The number of particles
per comoving volume, $R^3n$, is actually proportional
to the ratio of the particle number density to the entropy density,
$n/s$, where $s \propto g_*T^3$ and $g_*$ counts
the effective number of ultrarelativistic degrees
of freedom.  So long as $g_*$ is constant, $s$ and
$n_\gamma$ are related by a constant numerical factor,
today about $7.04$.}  of order unity for
temperatures $T\gg m/3$; of order $(m/T)^{3/2}\exp (-m/T)$
for $T\ll m/3$.  For reference, the fraction of critical
density contributed by a relic species is
\begin{equation}
\Omega h^2 \simeq \left({m\over 25\eV}\right)\, \left({n\over n_\gamma}
\right) .
\end{equation}
If equilibrium were the entire story, relic abundances
would be far too small to be of any interest.

Consider, a stable, massive particle species;
its abundance is necessarily regulated by
annihilations and pair creations.  In the expanding
Universe the temperature is decreasing, $\dot T/T \simeq -H$;
equilibrium can be maintained only
if annihilations and pair creations occur rapidly
on the expansion timescale, $H^{-1}$.  Because of
the temperature dependence of equilibrium number
densities and of cross sections, annihilation and
pair creation reactions eventually become ineffective (``freeze out'')
and the abundance of a particle species relative
to photons approaches a constant value (``freezes in'')
\cite{freeze}.

If freeze out occurs when the species is relativistic,
then the species' relic abundance is comparable to
that of photons.  Such a species is referred to
as a hot relic; a light (mass $\la \MeV$)
neutrino species is a hot relic.

On the other hand, if freeze out occurs when
the species is nonrelativistic, then its
relic abundance is significantly less than
that of photons, and depends inversely
upon its annihilation
cross section (in thermal equilibrium the annihilation
rate and pair creation rate are related by detailed
balance).  The relic abundance is
\begin{equation}
\left({n\over n_\gamma}\right) \sim
{ \ln (0.01 m\mpl \langle\sigma v\rangle_{\rm ann})
\over m \mpl \langle\sigma v\rangle_{\rm ann}}\qquad \Rightarrow
\qquad \Omega \sim {10^{-3} \over T_0\mpl
\langle\sigma v\rangle_{\rm ann}};
\end{equation}
where the second relation follows from the fact
that $\rho_{\rm CRIT}\sim 10^4T_0^4$.
This formula is quite remarkable:  Neglecting the
logarithmic factor and the overall numerical constant,
it implies that the fraction of critical density
contributed by a cold relic only depends upon its
annihilation cross section, and, further, that $\Omega \sim 1$
obtains for $\langle\sigma v\rangle_{\rm ann} \sim
10^{-3}/T_0\mpl \sim 10^{-37}\rcm^2$!
This is very roughly a weak-interaction
cross section ($\approx \GeV^2 G_F^2$), and
indicates that a stable particle with weak interactions
will necessarily have a relic abundance comparable to the critical
density.  A stable neutrino of mass a few GeV would
fit the bill were it not ruled out by experiment \cite{lw}.
The neutralino fits the bill nicely, as its interactions
with ordinary matter are roughly weak.

The final case is warm dark matter.  If
a species decouples while it is still relativistic,
but very early on ($T\gg 1\GeV$), then after it decouples its abundance
relative to photons will be diminished as various species
disappear and transfer their entropy to the photons
(and other species).  In this case, its abundance is
less than that of photons, but not exponentially less,
and so closure density obtains for masses in the $\keV$
range; plausible warm dark matter candidates include
the axino \cite{axino} and a light gravitino \cite{gravitino}.
(This dilution by ``entropy transfer''
is precisely what makes the relic neutrino
temperature and abundance less than that of photons.)

\subsection{Skew relics}

Implicit in the previous discussion is the assumption
that the particle and its antiparticle were equally abundant.
If there is an asymmetry between particle and antiparticle
and net particle number
is conserved, then the relic abundance can become
no smaller than the net particle number per photon \cite{skew}.
Provided that annihilations can reduce the particle's abundance
to this level, the relic abundance is determined by the
particle-antiparticle asymmetry.

Baryons are an example of a skew relic;
were it not for the asymmetry between baryons and
antibaryons, the relic abundance of each would
be about $10^{-18}$ that of photons \cite{steigman}.  The mass
density contributed by a skew relic is
\begin{equation}
\Omega_X h^2 \sim \left({\eta_X \over 10^{-10}}\right)
\,\left({m_X \over 250\GeV}\right) ;
\end{equation}
where $\eta_X$ is the particle-antiparticle asymmetry
relative to photons.  A stable neutrino
species with mass of order $100\GeV$ and asymmetry
of order the baryon asymmetry could provide closure
density.  (Neither the precision measurements of
the width of the $Z^0$ boson nor nucleosynthesis
preclude such a fourth neutrino species;
dark matter searches employing ionization
detectors do unless the mass exceeds
a $\TeV$ or so \cite{ion}.)

\subsection{Nonthermal relics}

The magnetic monopole and axion are examples of particles
whose relic abundance involves coherent, nonthermal processes.
Monopoles are produced as (point-like topological)
defects in the GUT symmetry breaking phase transition \cite{mono}.
On the basis of causality
considerations one expects of the order of one
monopole per horizon volume (at the time of the
phase transition), which leads to a relic abundance
of order $n/n_\gamma \sim (T/\mpl )^3$.  For the GUT
phase transition, $T\sim 10^{15}\GeV$ or so, which results
in a gross overabundance of monopoles (very crudely,
``$\Omega \sim 10^{12}$'').  This is the monopole
problem.  Inflation can solve the monopole problem
provided that the GUT phase transition occurs before
inflation, so that monopoles are diluted by the massive
entropy production.  This being said, it appears that
monopoles are a terrible dark matter candidate;
however, scenarios have been proposed where their relic
abundance can be close to critical \cite{mono}.

Axions arise
not only as thermal relics, but also due to two nonthermal
processes, the misalignment process and the
decay of axionic strings \cite{axlimits}.
For the interesting axion masses, $10^{-6}\eV$
to $10^{-4}\eV$, their thermal relic abundance cannot
come close to closure density.  Since there is
some disagreement as to the importance of the
axionic-string decay process \cite{axstring} and it is impotent
in an inflationary Universe I will focus
on the misalignment process \cite{axion}.

It is the $\Theta$ parameter of QCD that leads
to the strong-$CP$ problem; $\Theta$ is an angular
parameter that controls the strength of the
offending instanton effects.  In the PQ solution
$\Theta$ becomes a dynamical variable whose
value is anchored at the $CP$-conserving value
of zero by the instanton effects themselves.  However, at
temperatures much greater than $1\GeV$ these
effects are impotent and the value of $\Theta$ is
left undetermined by dynamical considerations.
Thus, one expects the value of $\Theta$ to be randomly
distributed in different causally independent
regions of the Universe.  When the QCD instanton
effects do become important $\Theta$ will in general
be ``misaligned''---i.e., not at $\Theta =0$---and
will evolve to toward $\Theta =0$; as it does,
$\Theta$ overshoots and is left oscillating.
These cosmic harmonic oscillations correspond
to a condensate of very nonrelativistic axions,
whose relic density is roughly
\begin{equation}
\Omega h^2 \simeq \left( {m\over 10^{-5}\eV} \right)^{-1.2}.
\end{equation}
The energy associated with the misalignment
of $\Theta$ is converted into an enormous number
of axions, about $10^9\cmm3$ for $m_a=10^{-5}\eV$.

\subsection{Significant-other relics}

While our first interest is in elucidating the nature
of the ubiquitous dark matter, it is possible that
there are a number of particle relics in our midst.
Needless to say, a particle relic that contributes significantly
less than closure could still be interesting---both from
the point of view of cosmology and of particle physics---moreover, it
could be detectable.  The CBR provides such an example:
$\Omega_\gamma \sim 10^{-4}$.  Until it was ruled out
by a telescope search for its decays, an $\eV$-mass
axion provided another possibility \cite{tele}.   If Nature is
supersymmetric and the lightest supersymmetry particle
is stable, it is difficult to avoid a supersymmetric
relic that contributes less than about $10^{-3}$ of closure density.
Magnetic monopoles provide yet another example.
If the earliest history of the Universe is as
interesting as many think, there may be many
relics whose abundance is far from
critical, but are still potentially detectable.

\subsection{Truly exotic relics}

Other more complicated explanations for the dark-matter problem
involving early Universe relics have been suggested.
Two suggestions have been made that would reconcile
a flat Universe with the observational data that
the amount of matter that clusters contributes only
20\% or so of critical density:  a ``relic cosmological
constant'' and dark-matter that decays a modest red
shift into relativistic debris which necessarily
remains unclustered \cite{krauss}.
In either case, dynamical measurements of $\Omega_0$
would not reveal the unclustered energy density---vacuum
energy or relativistic particles---and would yield
values of order 20\%.  On the other hand, kinematic
measurements could reveal the presence of the unclustered
energy density \cite{jc}.  In either case, a new cosmic coincidence
comes into play:  a cosmological constant that becomes
dynamically important in the current epoch, or a particle
whose lifetime is comparable to the age of the Universe.

A relic cosmological constant provokes further discussion.
Historically, cosmologists have turned to the cosmological constant
when faced with a crisis.  In the context of quantum-field theory
it is actually the {\it absence} of an enormous ($\Lambda \sim
10^{122}G^{-1}$) cosmological
constant associated with the zero-point energy
of quantum fluctuations of the fundamental fields
that is a mystery.  To confuse the situation further,
several authors have argued that a
Universe with a cosmological constant, cold dark matter
and baryons is currently the best-fit Universe,
in terms of the age of the Universe, dynamical measurements
of $\Omega_0$, and the formation of structure \cite{bestfit}.

Other puzzles have motivated suggestions for ``specialized relics.''
Sciama and others have argued for an unstable neutrino
species whose radiative decays would lead to
efficient re-ionization of the Universe \cite{reion}.
Recently, ``cocktails'' of two particle relics---30\%
neutrinos and 70\% cold dark matter---have
been advocated to make the cold dark matter scenario
for structure formation better agree with observations \cite{mixed}.

\subsection{A new cosmic ratio}

If the bulk of the mass density is in the form
of nonbaryonic dark matter, then cosmologists---and
particle physicists---have a new dimensionless number
to explain:  The ratio of ordinary matter to
exotic matter.  Why it is of the order of
unity and not say $10^{-20}$ or $10^{20}$?
The value of this ratio has important consequences
for the evolution of the Universe, and the fact
that it is of order unity is at the heart of
many cosmological observations---e.g., the
the halo/disk conspiracy in rotation curves,
the stability of galactic disks, and even the
formation of stars.

While there is presently no good explanation for why
this ratio is of order unity, it necessarily
involves fundamental physics.  For example, consider
a skew relic whose asymmetry is comparable
to the baryon asymmetry; then the ratio is just that
of the exotic particle's mass to the mass of a baryon.
For other relics, requiring that this ratio
be of order unity implies special relationships
between fundamental energy scales in physics \cite{ratio}.

\section{Detection}

The nonbaryonic dark-matter hypothesis is a very
bold one---and fortunately it is testable.
While no cosmological experiment or observation is
easy, especially the search for a particle whose interactions
could be as different as those of an axion and a neutralino,
thanks to the creative efforts of many
there are manifold approaches to the problem of detection \cite{detect}.

First, there are the direct schemes, where the halo
dark-matter particles in our local neighborhood (density
$5\times 10^{-25}\gcmm3$) are sought out.
For axions, the approach is based upon a very clever
idea of Sikivie that takes advantage of the axion's
coupling to two photons \cite{sikivie}.  A microwave cavity is
immersed in a very strong magnetic field which causes
halo axions to be converted to photons and excite resonant modes
of the cavity; several ``proof of principle
experiments'' have been built and operated and a new
generation of Sikivie detectors with sufficient
sensitivity to detect halo axions are being built \cite{karl}.
Neutralino detectors exploit the neutralino's
roughly weak interactions with ordinary matter:  When
a multi-GeV mass neutralino scatters off a nucleus
it deposits an energy of order a keV.
The annihilation cross section and elastic
cross section are related by
``crossing'' and thus the scattering cross
section too should be of order $10^{-37}\rcm^2$; this implies
an event rate of the order of 1 per day per kg.  A new generation
of low-background, low-threshold cryogenic detectors are
being developed to search for neutralinos in our halo \cite{sado}.
While the magnetic monopole must considered a
long shot dark-matter candidate, a football-field
sized detector called MACRO is just coming on line
and will achieve a sensitivity of about
$10^{-16}\rcm^{-2}{\rm sr}^{-1} \sec^{-1}$ \cite{macro}.

Next, there are indirect searches, which involve seeking out
the decay or annihilation products of dark-matter particles.
For example, dark-matter annihilations in the halo
of our galaxy can produce high-energy positrons that
can be detected \cite{positron}.  The most promising idea involves
annihilations of dark matter particles that accumulate
in the sun and the earth \cite{annihilate};
the annihilation products include
high-energy neutrinos that can be detected
in large, underground earth-based detectors, such
as MACRO \cite{macro}.  A sizable portion of the neutralino
parameter space can be explored by searching for
high-energy neutrinos from the sun and the earth \cite{kamio}.

Finally, there are numerous laboratory and astrophysical
experiments that bear on the existence of particle
dark matter.  Searches for the supersymmetric partners
of the known particles are taking place at every
accelerator in the world; the discovery of even
one superpartner would not only provide strong evidence
for the existence of the neutralino, but would also
help to narrow the parameter space.  There are a host
of experiments that bear on the issue of neutrino
masses:  experiments designed to measure the
electron-neutrino mass; neutrino oscillation/mixing
experiments; solar neutrino experiments; and
searches for neutrinos from type II supernovae.

\section{Concluding Remarks}

The theorists' prejudice of a flat Universe dominated by
nonbaryonic dark matter is at present just that!
However, I hope that I have convinced the reader
that:  (1) the dark matter question is a most pressing
one which now involves both cosmologists and
particle physicists; (2) the theorists' prejudice
is well motivated by both theoretical and observational
considerations; and (3) most importantly, the
particle dark matter hypothesis can and is being tested.
While cosmological experiments are inherently difficult
and we cannot test every dark-matter
candidate, I am optimistic.
The most promising dark-matter candidates are
detectable and the dark-matter problem has attracted
the attention of many of the most talented
experimentalists from both cosmology and particle physics.
While this is no guarantee that we will have an
answer soon, what more could one ask?
And if that isn't enough, there is the payoff:
Identifying and quantifying the primary substance
of the Universe and discovering ``new physics'' in
the process!

\vskip 0.5in

This work was supported in part by the DOE (at
Fermilab and Chicago) and by the NASA (through
NAGW-2381 at Fermilab).

\end{document}